\documentclass[epj,final]{svjour}
\usepackage{graphicx}% Include figure files
\usepackage{dcolumn}% Align table columns on decimal point
\usepackage{bm}% bold math
\usepackage{subfig}
\usepackage{amsmath,amssymb}
\usepackage{epstopdf}
\usepackage{textcomp}

\newcommand{\Eqref}[1]{Eq.~(\ref{#1})}

\newcommand{\nn}{\nonumber}
\newcommand{\be}{\begin{equation}}
\newcommand{\ee}{\end{equation}}
\newcommand{\alf}{{Alfv\'en~}}
\newcommand{\bear}{\begin{eqnarray}}
\newcommand{\eear}{\end{eqnarray}}
\newcommand{\Va}{V_\mathrm A}

\def \openone{\hbox{$1\hskip -1.2pt\vrule depth 0pt height 1.6ex width 0.7pt\vrule depth 0pt height 0.3pt width 0.12em$}}

\begin{document}

%\preprint{copy for friends -- not for distribution, corona20oct.tex}
\title{Heating of the solar corona by Alfv\'en waves -- self-induced opacity}
\author{\hspace{1.1cm}T.~M.~Mishonov \and N.~I.~Zahariev \and R.~V.~Topchiyska \and B.~V.~Lazov \and S.~B.~Mladenov \and A.~M.~Varonov}
\institute{\center{Department of Theoretical Physics, Faculty of Physics, St.~Clement of Ohrid University at Sofia,\\ 5 J. Bourchier blvd., BG-1164 Sofia, Bulgaria}\\[3pt]
\email{tmishonov@phys.uni-sofia.bg, nedeltcho.zahariev@gmail.com, rtopchiyska@gmail.com,\\ boian\_lazov@phys.uni-sofia.bg, \mailname{smladenov@phys.uni-sofia.bg}, avaronov@phys.uni-sofia.bg}}
\date{}

\abstract{Static distributions of temperature and wind velocity at the transition region are calculated within the framework of magnetohydrodynamics (MHD) of completely ionized hydrogen plasma. The numerical solution of the derived equations gives the width of the transition layer between the chromosphere and the corona as a self-induced opacity of high-frequency Alfv\'en waves (AW). The domain wall is direct consequence of the self-consistent MHD treatment of AW propagation. The low-frequency MHD waves coming from the Sun are strongly reflected by the narrow transition layer, while the high-frequency waves are absorbed -- that is why we predict considerable spectral density of the AW in the photosphere. The numerical method allows consideration of incoming AW with arbitrary spectral density. The idea that Alfv\'en waves might heat the solar corona belongs to Alfv\'en, we simply solved the corresponding MHD equations. The comparison of the solution to the experiment is crucial for revealing the heating mechanism.
}
\maketitle
%\keywords{Alfv\'en waves -- Magnetohydrodynamics -- Plasma -- Numerical Methods -- Solar corona}

%%%%%%%%%%%%%%%%%%%%%%%%%%%%%%%%%%%%%%%%%%%%%%%%%%%%%%%%%
\section{\alf model for corona heating}
%%%%%%%%%%%%%%%%%%%%%%%%%%%%%%%%%%%%%%%%%%%%%%%%%%%%%%%%%
The discovery of the lines of the multiply ionized iron in the solar corona spectrum \cite{Swings:43} posed an important problem for the fundamental physics - what is the mechanism of the heating of the solar corona and why the temperature of the corona is 100 times larger then the temperature of the photosphere.

The first idea by \alf \cite{Alfven:47} was that \alf waves (AW) \cite{Alfven:42} are the mechanism for heating the corona. AW are generated by the turbulence in the convection zone and propagate along the magnetic field lines. Absorption is proportional to $\omega^2$ and the heating comes from high-frequency AW. Alfv\'en's idea for the viscous heating of plasma by absorption of AW was analyzed in the theoretical work by Heyvaerts \cite{Heyvaerts:83}. In support of this idea is the work by Chitta \cite{Chitta:12} (Figures 8, 9 therein). The authors came to the conclusion that the spectral density of AW satisfies a power law with an index of 1.59. This gives a strong hint that this scaling can be extrapolated in the nearest spectral range for times less than 1~s and frequencies in the Hz range. Furthermore in the work by Tomczyk \cite{Tomczyk:09} it is stated that there exist very few direct measurements of the strength and orientation of coronal magnetic fields, meaning that the mechanisms responsible for heating the corona, driving the solar wind, and initiating coronal mass ejections remain poorly understood. After the launch of Hinode, however, the well-forgotten spatially and temporally ubiquitous waves in the solar corona \cite{Tomczyk:07} came again into the limelight and gave strong support for the idea of Alfv\'en. A clear presence of outward and inward propagating waves in the corona was noted. $k-\omega$ diagnostics revealed coronal wave power spectrum with an exponent of $\approx -\frac{3}{2}$ (cf. Fig. 2 of \cite{Tomczyk:09}). %The low frequency AW, on the other hand, reach the Earth orbit and thanks to the magnetometers on the various satellites we ``hear'' the basses of the great symphony of solar turbulence.

The observational data for the temperature profile of the solar corona show that the transition layer is extremely thin compared to the radius of the Sun \cite{Aschwanden:05}. The width of the transition layer $\lambda$ may be evaluated using the logarithmic derivative of the temperature, $\lambda=\mathrm{max}(\frac{\mathrm{d}T}{T\mathrm{d}x})$. In order to qualitatively explain this small width, in \cite{Mishonov:07} the idea of self-induced opacity of the plasma for Alfv\'en waves was introduced. A similar idea was analysed also by Suzuki \cite{Suzuki:08} (see also \cite{Ofman:10,Ofman:05,Nakariakov:00,Ofman:98,Ofman:95}). In the current paper we give a numerical realisation of this idea by calculating the width of the transition layer using the framework of MHD. Without a doubt AW should be present in the solar corona in the form of the torsional nonlinear cascade \cite{Farahani:12}.

The purpose of the present work is to examine whether the initial Alfv\'en idea is correct and to solve the MHD equations which give the dependence of the temperature on the height  $T(x)$ and the related velocity of the solar wind $U(x)$ supposing static density of the incoming AW. We illustrate Alfv\'en's idea by an MHD calculation for completely ionized hydrogen plasma. Due to the high density of the transition layer MHD is an adequate tool to analyze the beginning of the process. Without a doubt the kinetic approach is indispensable for the treatment of low density solar corona but this problem is beyond the purpose of the present work.

Our starting point are the MHD equations for the velocity field $\mathbf{v}$ and magnetic field $\mathbf{B}$
\bear
	&&\partial_t\rho+\mathrm{div}\mathbf{j}=0,  \qquad \mathbf{j}=\rho \mathbf{v},								\label{conslaws}\\
	&&\partial_t\left(\frac{1}{2}\rho\mathbf v^2+\varepsilon+\frac{\mathbf  B^2}{2\mu_0} \right)+\mathrm{div}\, \mathbf q=0		\label{conslaws2},\\
	&&\partial_t(\rho \mathbf v)+\nabla\cdot \mathbf{\Pi}=0,												\label{conslaws3}
\eear
where
\be
	\mathbf q=\rho\left(\frac{1}{2}\mathbf{v}^2+h\right)\mathbf{v}+\mathbf{v}\cdot\mathbf{\Pi}^\mathrm{(visc)}-\varkappa\nabla T+\mathbf{S}
	\label{vec_q}
\ee
is the energy density flux, $\rho$ is the mass density, $\varepsilon$ is the internal energy density, $\varkappa$ is the thermal conductivity, $h$ is the enthalpy per unit mass;
\be
	\mathbf S=\frac1{\mu_0}\left[\mathbf{B}\times\left(\mathbf{v}\times\mathbf{B}\right)-\nu_\mathrm{m} \mathbf{B}\times\mathrm{rot}\mathbf{B}\right],
\ee
is the Poynting vector and $\nu_\mathrm{m}\equiv c^2\varepsilon_0\varrho$ is the magnetic diffusivity determined by Ohmic resistance $\varrho$ and vacuum susceptibility $\varepsilon_0$; vacuum permeability is $\mu_0.$ For hot enough plasma $\varrho$ is negligible. The total momentum flux
\be
	\mathbf{\Pi}=\rho \mathbf{vv}+P\openone+\mathbf\Pi^\mathrm{(visc)}+\mathbf\Pi^\mathrm{(Maxw)}
	\label{Pi_xx}
\ee
is a sum of the inviscid part $\rho \mathbf{vv}+P$ of the fluid, with pressure $P$,
\be
	\Pi_{ik}^\mathrm{(visc)}=-\eta\left(\partial_i v_k+\partial_k v_i-\frac{2}{3}\delta_{ik}\nabla\cdot \mathbf v\right)-\zeta\delta_{ik}\nabla\cdot\mathbf v,
\ee
the viscous part of the stress tensor, with viscosity $\eta$ and second viscosity $\zeta,$ and lastly, the Maxwell stress tensor
\be
	-\Pi_{ik}^\mathrm{(Maxw)}=\frac1{\mu_0}\left(B_iB_k-\frac{1}{2}\mathbf B^2\delta_{ik}\right),
\ee
with $\delta_{ik}$ the Kronecker delta. We model coronal plasma with completely ionized hydrogen plasma with the following parameters:
\bear
	\label{kappa_eta}
	&&\varkappa=0.9\frac{T^{5/2}}{e^4m_e^{1/2}\Lambda},\quad\eta=0.4\frac{m_p^{1/2}T^{5/2}}{e^4\Lambda},\quad\zeta\approx 0,\\
	&&\Lambda=\ln\left(\frac{r_\mathrm{D}T}{e^2}\right),\quad\frac1{r_\mathrm{D}^2}=\frac{4\pi e^2n_\mathrm{tot}}T,\quad e^2\equiv\frac{q_e^2}{4\pi\varepsilon_0}\nn
\eear
where $q_e$ is the electron charge, $m_e$ is the mass of electron, $m_p$ is the proton mass, $T$ is the temperature and $n_\mathrm{tot}=n_e+n_p$ is the total density of electrons and protons; $\rho=m_pn_p.$ We suppose that $\mu_0=4\pi$ and $\varepsilon_0=1/4\pi$, but in the practical system all formulae are the same; as well as in Heaviside-Lorentz units where $\mu_0=1$ and $\varepsilon_0=1$. As we mentioned above
\be
	\label{nu_m_k}
	\nu_\mathrm{m}=\frac{c^2}{4\pi}\frac{e^2m_e^{1/2}\Lambda}{0.6\,T^{3/2}}\ll\nu_\mathrm{k}\equiv\frac{\eta}{\rho}=\frac{0.4\,T^{5/2}}{e^4m_p^{1/2}n_p\Lambda};
\ee
i.e. the hot hydrogen plasma is sticky, dilute, and ``superconducting''. Here $\nu_{\mathrm{k}}$ is the kinematic viscosity. We can introduce the magnetic Prandtl number as the ratio between the kinematic and the magnetic viscosity,
\be 
\mathrm{Pr}_\mathrm{m}=\frac{\nu_\mathrm{k}}{\nu_\mathrm{m}}=\frac{0.96\pi}{e^6c^2\sqrt{m_pm_e}n_p\Lambda^2(T)}T^4.
\ee
In terms of this ratio the heating of the solar corona by AW is an MHD problem with high Prandtl numbers $\mathrm{Pr}_\mathrm{m}\gg 1$ in the whole region, where the energy of the waves is transformed into energy of the hot solar wind. While for the cold chromosphere the magnetic Prandtl number is of order of one, for the hot corona with a 100 times higher $T$ it increases by 8 orders of magnitude. For this reason, in the whole interval where we consider MHD heating, the coefficient $\nu_\mathrm{m}$ is negligible. In other words, the formal inclusion of Ohmic heating in the MHD equations will not change the profile $T(x)$ in the range of highest temperatures which is exactly the purpose of our consideration. Therefore, for illustrative purposes, we can take $\nu_\mathrm{m}=0$. The coronal heating mechanism can be revealed without additional accessories. Let us mention also the relations $\varkappa\varrho=1.5\,T/q_e^2$ and $\eta/\varkappa\approx\frac49\sqrt{m_em_p}$ leading to
\be
	\varrho=\frac1{4\pi \varepsilon_0}\frac{e^2 m_e^{1/2}{\Lambda}}{0.6 \,T^{3/2}}.
\ee

The applicability of our equations is governed by an additional condition:
\be
\nu_{pp}=\frac{4\pi e^2n\Lambda}{T^{3/2}\sqrt{m_p}}\gg\omega_{\mathrm{c}p}=\frac{eB}{m_{p}}.
\label{eq:crit}
\ee
If in the dense plasma the frequency of the proton-proton collisions is much smaller than the proton cyclotron frequency, the magnetic field can be considered a perturbation and for the kinetic coefficients we can use the zero magnetic field formulas.

Lastly we need to mention the existence of radiative losses. However, the thin transition layer has width much smaller not only than Earth's orbit radius but than Earth's radius and thus these losses are neglectable when considering the mechanism for coronal heating. The hot corona exists in broad ranges where the heating mechanism cannot be effective. For this reason, in the narrow range of the transition region the radiation losses are negligible compared to the intensive heating, no matter what the concrete mechanism is. Quantitatively, this means that radiation power per unit volume $P_\mathrm{rad}\ll nTw/v$, where $n$ and $T$ are the number density and temperature of the corona, $w$ is the width of the transition region, and $v$ is the velocity of the solar wind.
%%%%%%%%%%%%%%%%%%%%%%%%%%%%%%%%%%%%%%%%%%%%%%%%%%%%%%%%%%
\section{MHD equations and energy fluxes}%%%%%%%%%%%%%%%%%%%%%%%%%%%%%%%%%%%%%%%%%%%
%%%%%%%%%%%%%%%%%%%%%%%%%%%%%%%%%%%%%%%%%%%%%%%%%%%%%%%%%%
The time derivative $\partial_t \mathrm{B}$ which implicitly participates in the energy conservation \Eqref{conslaws2} at zero Ohmic resistivity obeys the equation
\bear
	\label{d_tB}
	\mathrm{d}_t\mathbf{B}&=&\mathbf{B}\cdot\nabla \mathbf{v}-\mathbf{B}\,\mathrm{div}\,\mathbf{v}+\mathrm{rot}\left(\nu_\mathrm{m}\mathrm{rot}\mathbf{B}\right),\\ \mathrm{d}_t&\equiv&\partial_t + \mathbf{v}\cdot\nabla\,\nonumber.
\eear
Analogously the momentum equation \Eqref{conslaws3} can be rewritten by the substantial derivative
\bear
	\rho\,\mathrm{d}_tv_i	&=	&-\partial_i P+\partial_k\left\{\eta\left(\partial_kv_i+\partial_iv_k-\frac23\delta_{ik}\partial_jv_j\right)\right\}\nn\\
					&	&+\partial_i\left(\zeta\partial_jv_j\right)-\frac1{\mu_0}\left(\mathbf{B}\times\mathrm{rot}\,\mathbf{B}\right)_i\,.
	\label{momentum}
\eear
In our model we consider AW propagating along magnetic field lines $\mathbf{B}_0.$ We focus our attention on the narrow transition layer, where the static magnetic field is almost homogeneous and the waves are within acceptable accuracy one dimensional. For the velocity and magnetic fields we assume
\bear
	\mathbf{v}(t,x)\!	&=	&\!U(x)\hat {\mathrm x}+u(t,x)\hat {\mathrm z},\nn\\
	\mathbf{B}(t,x)\!	&=	&\!B_0\hat {\mathrm x}+ b(t,x)\hat{\mathrm  z},
\eear
with homogeneous magnetic field $B_0\hat{\mathrm x}$ perpendicular to the surface of the Sun. The transverse wave amplitudes of the velocity $u(t,x)$ and magnetic field $b(t,x)$ we represent with the Fourier integrals
\bear
	u(t,x)&=&\int_{-\infty}^\infty \tilde u(\omega,x)\mathrm{e}^{-\mathrm i \omega t}\frac{\mathrm{d}\omega}{2\pi},\\
	b(t,x)&=&\int_{-\infty}^\infty \tilde b(\omega,x)\mathrm{e}^{-\mathrm i \omega t}\frac{\mathrm{d}\omega}{2\pi}.
\eear
For illustrative purposes it is convenient to consider \hyphenation{mono-chro-matic}monochromatic AW with $u(t,x)=\hat u(x) \mathrm{e}^{-\mathrm i \omega t}$ and $b(t,x)=\hat b(x) \mathrm{e}^{-\mathrm i \omega t}$.
%------------------------------------------------------------------------------------------------------------------------------------------------------------
\subsection{Wave equations}
%------------------------------------------------------------------------------------------------------------------------------------------------------------
For linearized waves the general MHD equations \Eqref{momentum} and \Eqref{d_tB} give the following system for $\hat u(x)$ and $\hat{\overline{b}}(x)\equiv\hat b(x)/B_0$
\bear
	\label{wave u}
	&&\left(-\mathrm{i}\omega+U\mathrm{d}_x\right)\hat u=V_\mathrm A^2\mathrm{d}_x\hat{\overline{b}}+\frac1\rho\mathrm{d}_x\left(\eta\mathrm{d}_x\hat u\right),\\
	&&-\mathrm i\omega\hat{\overline{b}}=\mathrm{d}_x\hat u -\mathrm{d}_x(U\hat{\overline{b}})+\mathrm{d}_x(\nu_\mathrm{m}\mathrm{d}_x\hat{\overline{b}}),		\label{wave beta}
\eear
where
\be
	\label{Va}
	\Va(x)=B_0/\sqrt{\mu_0\rho(x)}
\ee
is the \alf velocity. In our numerical analysis we solve the first order linear system of equations
\bear
	\label{wave_equation}
	-\mathrm i\mathrm{d}_x 	&\Psi&=\mathsf K \Psi,\\
					&\Psi&\equiv\left(\begin{array}{c}
		\hat u\\
		\hat {\overline{b}}\\
		\hat w
	\end{array}\right),\quad\mathsf K=\frac{\mathrm i}{\nu_{\mathrm k}U}\mathsf M,\nn\\
					&\Psi^\dagger&=\left(\hat u^*, \hat {\overline{b}}^*, \hat w^*\right),\nn
\eear
where $ \hat w\equiv\mathrm{d}_x\hat u,$ and
\be
	\mathsf{M}\!\equiv\!\!\left(\begin{array}{ccc}
		0				&0									&-\nu_{\mathrm k} U\\
		0				&\nu_{\mathrm k}(-\mathrm{i}\omega\!+\mathrm{d}_xU)	&-\nu_{\mathrm k}\\
		\mathrm{i}\omega	 U	&-V_\mathrm A^2(-\mathrm{i}\omega\!+\mathrm{d}_xU)	&\left(V_\mathrm A^2\!-\!U^2\right)\!+\!\frac U\rho\mathrm{d}_x\eta
	\end{array}\right)\!.\nn
\ee
For homogeneous medium with constant $\eta,$ $\rho,$ $\Va$, and $U$, in short for constant wave-vector matrix $\mathsf{K},$ the exponential substitution $\Psi\propto \exp(\mathrm ikx)$ in \Eqref{wave_equation} or equivalently \Eqref{wave u} and \Eqref{wave beta} gives the secular equation
\bear
	\label{secular}
	&&\mathrm i\,\nu_{\mathrm k}U\mathrm{det}\left(\mathsf{K}-k\openone\right)\\
	&&\qquad =\omega_\mathrm{D}\left(\omega_\mathrm{D}+\mathrm{i}\nu_{\mathrm k}k^2\right)-\Va^2k^2=0,\nn
\eear
where $\omega_\mathrm D\equiv \omega-k U$ is the Doppler shifted frequency.  This secular equation gives the well-known dispersion relation $$\omega_\mathrm D\left(\omega_\mathrm D +\mathrm i\nu_\mathrm k k^2\right) =\Va^2k^2$$ of the AW. This equation is quadratic with respect to $\omega$ and cubic with respect to $k.$ If the Ohmic resistance is taken into account, the secular equation takes the form
\be
\left(\omega_\mathrm{D}+i\nu_\mathrm{m}k^2\right)\left(\omega_\mathrm{D}+\mathrm{i}\nu_{\mathrm k}k^2\right)=\Va^2k^2.
\ee
%------------------------------------------------------------------------------------------------------------------------------------------------------------
\subsection{Wind variables}
%------------------------------------------------------------------------------------------------------------------------------------------------------------
We solve the wave equation \Eqref{wave_equation} from ``Sun' surface'' $x=0$ to some distance large enough $x=l$, where the short wavelength AW are almost absorbed. This distance is much bigger than the \emph{width of the transition layer } $\lambda$, but much smaller than solar radius. The considered one-dimensional $0<x<l$ time-independent problem has three integrals corresponding to the three conservation laws related to mass, energy and momentum. The mass conservation law \Eqref{conslaws} gives the constant flow
\be
	\label{mass_conservation}
	j=\rho(x)U(x)=\rho_0 U_0=\rho_l U_l=\mathrm{const},
\ee
where $\rho_0=\rho(0),$ $\rho_l=\rho(l),$ $U_0=U(0),$ and $U_l=U(l).$ The energy conservation law reduces to a constant flux along the $x$-axis
\be
	\label{energy_flux}
	q=q_\mathrm{wind}^\mathrm{ideal}(x)+\tilde q(x)=\rho U\left(\frac12U^2+h\right)+\tilde q=\mbox{const.}
\ee
Here the first term describes the energy of the ideal wind, i.e. a wind from an ideal (inviscid) fluid. The second term $\tilde q(x)$ includes all other energy fluxes; in our notations tilde will denote sum of the non-ideal (dissipative) terms of the wind and wave terms. In detail the non-ideal part of the energy flux $\tilde q(x)$ consists of: the wave kinetic energy $\propto \left|\hat u\right|^2$, viscosity (wind $\propto \frac43\eta+\zeta$ and wave $\propto \eta$ components), heat conductivity $\propto \varkappa$, and Poynting vector $\propto \hat b^*,$
\bear
	\tilde q(x)	&\equiv	&\frac{j}{4}\left|\hat u\right|^2-\xi U\mathrm{d}_x U-\frac14 \eta\,\mathrm d_x \! \left|\hat u\right|^2-\varkappa \, \mathrm{d}_x\!T \nn\\
			&		&+\frac1{2\mu_0} \left(U\left|\hat b \right|^2\!-B_0\,\mathrm{Re}(\hat b^*\hat u)\right),
	\label{non-ideal_energy_flux}
\eear
where $\xi\equiv \frac43\eta+\zeta.$ Here time averaged energy flux is represented by the amplitudes of the monochromatic oscillations, this is a standard procedure for alternating current processes. In our case we have, for example, $\left<\hat u^2\right>_t=\left<\left(\mathrm {Re}\,\hat u\right)^2\right>_t=\left<\frac14(\hat u+\hat u^*)^2\right>_t=\frac12\left|\hat u\right|^2.$ The other terms from \Eqref{vec_q} are averaged in a similar way like in the equation above.

The momentum conservation law \Eqref{Pi_xx} gives constant flux $\Pi=\Pi_{xx}$
\be
	\label{momentum_flux}
	\Pi=\Pi_\mathrm{wind}^\mathrm{ideal}(x)+\tilde\Pi(x)=\rho UU +P+ \tilde\Pi,
\ee
the sum of the ideal wind fluid and the other terms
\be
	\label{Non-Ideal_Momentum}
	\tilde\Pi(x)\equiv\frac{1}{4\mu_0}\left|\,\hat b \right|^2-\xi\mathrm{d}_x U,
\ee
which take into account the wave part of the Maxwell stress tensor $\propto b^2$ and viscosity of the wind $\propto \xi$.

We have to solve the hydrodynamic problem for calculation of wind velocity and temperature at known energy and momentum fluxes. The problem is formally reduced to analogous one for a jet engine, cf. Ref. \cite{Feynman:65}. We approximate the corona as completely ionized hydrogen plasma, i.e. electrically neutral mixture of electrons and protons. The experimental data tells us that proton temperature $T_p$ is higher than electron one $T_e.$ This is an important hint that heating goes through the viscosity determined mainly by protons. However for illustration purpose and simplicity we assume proton and electron temperatures to be equal $T_e=T_p=T.$ For such an ideal (in thermodynamic sense) gas the local sound velocity is
\bear
	&&c_\mathrm{s}^2(x)=\frac{c_p}{c_v}\frac{P}{\rho}=\gamma\frac{T}{\left<m\right>},\quad\gamma=\frac{c_p}{c_v}=\frac53,\\
	&&\left<m\right>=\frac{n_pm_p+n_em_e}{n_p+n_e}\approx \frac12m_p,\quad n_e=n_p=\frac{1}{2}n_\mathrm{tot},\nn\\
	&&P=n_\mathrm{tot}T=\frac{\rho T}{\left<m\right>}=\frac{j}{U}\frac{T}{\left<m\right>},\quad h=c_p\frac{T}{\left<m\right>}=\frac{\varepsilon+P}{\rho},\nn
\eear
where, as we mentioned earlier, $h$ is the enthalpy per unit mass and $\varepsilon$ is the density of internal energy. Although there are some hints for different values of the adiabatic index $\gamma$ \cite{Doorsselaere:11}, we will use the traditional value of 5/3 for our calculations since this choice will not change the essence of our presentation.

In order to alleviate the final formulae we introduce two dimensionless variables $\chi$ and $\tau$ which represent the non-ideal part of the energy and momentum flux respectively
\be
	\label{sigma-tau}
	\chi(x)\equiv\left.\frac{\tilde{q}(x)}{\rho_0^{} U_0^3}\right|_x^0,\quad\tau(x)\equiv\left.\frac{\tilde{\Pi}(x)}{\rho_0^{} U_0^2}\right|_x^0.
\ee
Here we could include gravitational field, Bremsstrahlung or other accessories, which are negligible for the narrow transition layer. Analogously, for the wind velocity and temperature we have
\be
	\overline{U}(x)\equiv\frac{U(x)}{U_0},\quad\Theta(x)\equiv\frac{T(x)}{\left< m\right>U_0^2},\quad\Theta_0=\Theta(0),\label{dlesswind}
\ee
where $U_0=U(0).$ The energy and momentum constant fluxes \Eqref{energy_flux} and \Eqref{momentum_flux} in the new notation take the form
\bear
	\frac {q-\tilde q(0)}{\rho_0U_0^3}&=&\frac12 \overline{U}^2+c_p\Theta-\chi=\frac12 +c_p\Theta_0,	\label{q/j}\\
	\frac {\Pi-\tilde\Pi(0)}{\rho_0U_0^2} &=& \overline{U}+\Theta/\overline{U}-\tau=1+\Theta_0.		\label{Pi/j}
\eear
From the second equation we express the dimensionless temperature $\Theta$ and substitute in the first one. Solving the quadratic equation for the wind velocity $U$ we derive
\be
	\label{U(x)}
	U=U_0\overline{U},\quad\overline U(x)=\frac{1}{\gamma+1}\left(\gamma+s^2+\gamma\tau(x)-\sqrt{\mathcal{D}(x)}\right),
\ee
where for the discriminant we have
\bear
	&&\!\!\!\mathcal{D}={(s^2-1)^2-2\chi(\gamma^2-1)+\gamma\tau\left[\gamma\tau+2\left(\gamma+s^2\right)\right]},\nn\\
	&&\!\!\!s^2\equiv \frac{c_\mathrm{s}^2(0)}{U_0^2}= \gamma\Theta_0,\quad c_\mathrm{s}^2(x)=\left(\frac {\partial P}{\partial \rho}\right)_{\!\!S}=\frac{\gamma T(x)}{\left< m \right>}.
\eear
If $\chi=0$ and $\tau=0$, we get the initial condition $\overline{U}|_{\chi=0,\tau=0}=1$. This condition determines the sign in front of $\sqrt{\mathcal{D}}$ in Eq. \eqref{U(x)}. Here $\gamma$ is the constant ratio of the heat capacities, and $s\equiv c_\mathrm s(0)/U_0$ is the ratio of the sound and wind velocity at $x=0$. We suppose that initial wind velocity is very small $U(0)\ll c_\mathrm s(0).$ The velocity distribution \Eqref{U(x)} can be substituted in \Eqref{Pi/j} and we derive the dimensionless equation for the temperature distribution
\bear
	\label{T(x)}
	&&T(x)=\left<m\right>U_0^2\Theta(x),\\
	&&\Theta(x)=\overline{U}(x)\left(1+\Theta_0+\tau(x)-\overline{U}(x)\right),\;\Theta_0=\frac{s^2}{\gamma}.\nn
\eear
The solutions for velocity $\overline U(x)$ \Eqref{U(x)} and temperature $\overline T(x)$ \Eqref{T(x)} distributions are important ingredients in our analysis and derivation of the self-consistent picture of the solar wind. We use a one dimensional approximation and in addition the constant flux of mass, energy and momentum gives 3 integrals of motion. This enables us to solve the nonlinear part of the problem analytically. That is why we do not solve the differential equations for the density $\rho(x)=\rho (0)U(0)/U(x)$, temperature $T(x)$ and wind velocity $U(x)$, and use analytical expressions containing the energy and momentum fluxes. Thus the numerical problem is reduced to a system of three linear differential equations.
%%%%%%%%%%%%%%%%%%%%%%%%%%%%%%%%%%%%%%%%%%%%%%%%%%%%%%
\subsection{Boundary conditions for the waves}%%%%%%%%%%%%%%%%%%%%%%%%%%%%%%%%%
%%%%%%%%%%%%%%%%%%%%%%%%%%%%%%%%%%%%%%%%%%%%%%%%%%%%%%
At known background wind variables $U(x)$ and $T(x)$ we can solve the wave equation \Eqref{wave_equation} for run-away AW at $x=l$. As we will see later the run-away boundary condition \Eqref{run-away} corresponds to right propagating AW at the right boundary of the interval. The wave equation \Eqref{wave_equation} is extremely stiff at
small viscosity, and numerical solution is possible to be obtained only downstream from $x=0$ to $x=l.$ We have to find the linear combination of left and right propagating waves at $x=0$, which gives the run-away condition at $x=l.$

The solution of wave equation according to \Eqref{energy_flux} determines the energy flux related to the propagation of AW
\bear
	\tilde q_\mathrm{wave}&&\!\!\!\left(\Psi(x)\right)\equiv\Psi^\dagger g\Psi=\frac{j}{4}\left|\hat u\right|^2-\frac12 \eta\, \mathrm{Re} \left(\hat u^*\hat w\right)\\
	&&-\frac{\nu_{\mathrm{m}}B_0^2}{2\mu_0}\mathrm{Re}\left(\hat {\overline{b}}^*\mathrm{d}_x \hat {\overline{b}}\right)+\frac{B_0^2}{2\mu_0} \left(U\left|\hat {\overline{b}} \right|^2\!\!-\mathrm{Re}\!\left(\hat {\overline{b}}^*\hat u\right)\right)\nn,
	\label{wave_energy_flux}
\eear
where
\be
	g(x)\equiv\left(\begin{array}{ccc}
		\frac14 j&-\frac{B_0^2}{4\mu_0}&-\frac14 \eta\\
		-\frac{B_0^2}{4\mu_0}&\frac{UB_0^2}{2\mu_0}&0\\
		-\frac14 \eta&0&0
	\end{array}\right).
\ee
Here $j$-term represents kinetic energy of the wave, $\eta$-term comes from the viscous part of the wave energy flux, and $B_0$-terms describe the Poynting vector of the wave. For illustrative purposes we consider hot enough plasma with negligible Ohmic resistance, i. e. $\nu_\mathrm{m}\approx 0$.

In order to take into account the boundary condition at $x=l$ we calculate the eigenvectors of the matrix $\mathsf{K}$, which according to \Eqref{wave_equation} determine the wave propagation in a homogeneous fluid with amplitude $\propto\exp(
\mathrm{i}k x).$ Then the eigenvalues of $\mathsf{K}$ give the complex wave-vectors
\be
	k=k^\prime+\mathrm{i}k^{\prime\prime}=\mbox{eigenvalue}(\mathsf{K}),
\ee
i.e.
\be
	\mathrm{det}\left(\mathsf{K}-k\openone\right)=0.
\ee
The three eigenvectors $\mathrm{L},$ $\mathrm D$ and $\mathrm R$ are ordered by spatial decrements of their eigenvalues
\be
	k^{\prime\prime}_\mathrm{L}<0<k^{\prime\prime}_\mathrm{R}< k^{\prime\prime}_\mathrm D,
\ee
and are normalized by the conditions
\be
	-\mathrm{L}^\dagger g\,\mathrm{L}=\mathrm{R}^\dagger g\,\mathrm{R}=\mathrm{D}^\dagger g\,\mathrm{D}=1,
\ee
where the sign corresponds to the direction of wave propagation. Notation L corresponds to left propagating wave, R to right propagating wave, and D for an overdamped at small viscosity mode.

For technical purposes we introduce the matrix notations
\be
	\mathrm{L}=\left(\begin{array}{c}
		\mathrm{L}_u(x)\\
		\mathrm{L}_{{b}}(x)\\
		\mathrm{L}_w(x)
	\end{array}\right),\quad \mathrm{R}=\left(\begin{array}{c}
		\mathrm{R}_u(x)\\
		\mathrm{R}_{{b}}(x)\\
		\mathrm{R}_w(x)
	\end{array}\right)\nn,
\ee
\be
	\mathrm{D}=\left(\begin{array}{c}
		\mathrm{D}_u(x)\\
		\mathrm{D}_{{b}}(x)\\
		\mathrm{D}_w(x)
	\end{array}\right).
\ee
For low enough frequencies $\omega\rightarrow 0$ and wind velocities the modes describe: 1) right-propagating AW with $k^{\prime}_\mathrm{R}\approx\omega/V_\mathrm{A}$ and small $k^{\prime\prime}_\mathrm{R}\approx\nu_\mathrm{k}\omega^2/2V_\mathrm{A}^3\ll k^{\prime}_\mathrm{R},$ 2) left propagating wave $k_\mathrm{L}=-k_\mathrm{R},$ and a diffusion overdamped mode $k^{\prime\prime}_\mathrm{D}\approx V_\mathrm{A}^2/\nu_\mathrm{k}U\gg k^{\prime}_\mathrm{D}$ which describes the drag of a static perturbation by the slow wind $U\ll V_\mathrm{A}$ in a fluid with small viscosity. In this low frequency and long wavelength limit the stiffness ratio of the eigenvalues is very large
\be
	r_{_\mathrm{DR}}=\frac{\left|k_\mathrm{D}\right|}{\left|k_\mathrm{R}\right|}\approx\frac{k^{\prime\prime}_\mathrm{D}}{k^{\prime}_\mathrm{R}}\approx\frac{V_\mathrm{A}^3}{\nu_\mathrm{k}U\omega}\gg1.
\ee
The strong inequality is applicable to the chromosphere where the viscosity of the cold plasma is very low. As we emphasized the wave equations \Eqref{wave_equation} form a very stiff system and indispensably has to be solved downstream from the
chromosphere $x=0$ to the corona $x=l$ using algorithms for stiff systems. Let
\be
	\psi_{_\mathrm{L}}(x)=\left(\begin{array}{c}
		u_{_\mathrm{L}}(x)\\
		{{b}}_{_\mathrm{L}}(x)\\
		w_{_\mathrm{L}}(x)
	\end{array}\right),\quad\psi_{_\mathrm{R}}(x)=\left(\begin{array}{c}
		u_{_\mathrm{R}}(x)\\
		{{b}}_{_\mathrm{R}}(x)\\
		w_{_\mathrm{R}}(x)
	\end{array}\right)
\ee
are the solutions of the wave equation \Eqref{wave_equation} with boundary conditions
\be
	\psi_{_\mathrm L}(0)=\mathrm{L}(0),\quad\psi_{_\mathrm R}(0)=\mathrm{R}(0).
\ee
We look for a solution as a linear combination
\be
	\label{psi-solution}
	\psi(x)=\psi_{_\mathrm R}(x)+r\,\psi_{_\mathrm L}(x),
\ee
in other words we suppose that from the low viscosity chromosphere plasma do not come overdamped
diffusion modes. The strong decay rate make them negligible at $x=0.$
Physically this means that AW (R-modes) are coming from the Sun
and some of them are reflected from the transition layer (L-modes)
\be
	\label{bound-cond-0}
	\psi(0)=\mathrm{R}(0)+r\,\mathrm{L}(0).
\ee

Analogously for the configuration of open corona we have to take into account the run-away boundary condition for which we suppose zero amplitude for the wave coming from infinity
\be
	\label{bound-cond-l}
	\psi(l)=\tilde t\,\mathrm{R}(l)+\tilde c\,\mathrm{D}(l).
\ee
Written by components
\be
	\left(\begin{array}{c}
		u_{_\mathrm{R}}(l)\\
		{{b}}_{_\mathrm{R}}(l)\\
		w_{_\mathrm{R}}(l)
	\end{array}\right)+r\left(\begin{array}{c}
		u_{_\mathrm{L}}(l)\\
		{{b}}_{_\mathrm{L}}(l)\\
		w_{_\mathrm{L}}(l)
	\end{array}\right)=\tilde t\left(\begin{array}{c}
		\mathrm R_u(l)\\
		\mathrm R_{{b}}(l)\\
		\mathrm R_w(l)
	\end{array}\right)+\tilde c\left(\begin{array}{c}
		\mathrm D_u(l)\\
		\mathrm D_{{b}}(l)\\
		\mathrm D_w(l)
	\end{array}\right).
\ee
This boundary condition gives a linear system of equation for the reflection coefficient $r$, transmission coefficient $\tilde t$ and the mode-conversion coefficient $\tilde c$.

For $l\rightarrow \infty$ when $\exp[-k_\mathrm{D}^{\prime\prime}(l)l]\ll1\exp[-k_\mathrm{R}^{\prime\prime}(l)l]$ the amplitude of D-mode is negligible and the run-away boundary condition reads
\be
	\label{run-away}
	\psi(l)=\psi_{_\mathrm R}(l)+r\psi_{_\mathrm L}(l)\approx \tilde t \mathrm{R}(l),
\ee
or by components
\be
	\left(\begin{array}{c}
		u_{_\mathrm{R}}(l)\\
		{\overline{b}}_{_\mathrm{R}}(l)
	\end{array}\right)+r\left(\begin{array}{c}
		u_{_\mathrm{L}}(l)\\
		{\overline{b}}_{_\mathrm{L}}(l)
	\end{array}\right)=\tilde t\left(\begin{array}{c}
		\mathrm R_u(l)\\
		\mathrm R_{\overline{b}}(l)
	\end{array}\right).
\ee
These systems give the amplitudes of the reflected wave $r$ and transmitted wave $\tilde t$ in the solution \Eqref{psi-solution}. For this solution we have the energy flux of transmitted $\mathcal{T}$ and reflected $\mathcal{R}$ waves
\bear
	&&\mathcal{T}\equiv\psi^\dagger(l)\,g(l)\,\psi(l)=\\
	&&\quad =|\tilde t|^2+|\tilde c|^2+\left(\tilde t\,\tilde c^*\mathrm{D}^\dagger(l)\,g(l)\,\mathrm{R}(l)+\mathrm{c.c.}\right),\nn\\
	&&1-\mathcal{R}\equiv\psi^\dagger(0)\,g(0)\,\psi(0)\\\nn
	&&\quad =1-|r|^2+\left(r^*\mathrm{L}^\dagger(0)\,g(0)\,\mathrm{R}(0)
+\mathrm{c.c.}\right).
\eear
Then we introduce the absorption coefficient
\be
\mathcal{A}\equiv -\psi^\dagger(x)\,g(x)\,\psi(x)\Big|_0^l=
1-\mathcal{R}-\mathcal{T}.%,
\ee
The described solution is normalized by unit energy flux of the R-wave.
If we wish to fix energy flux of the right propagating wave to be $q_\mathrm{wave}(0)$
we have to make the renormalization
\be
\label{norm_psi}
\Psi(x)=A_\mathrm{wave}\psi(x), %=\sqrt{q_\mathrm{wave(0)}}\,\psi(x)
\ee
using a parameter $A_{\mathrm{wave}}$.
In this section we have described Absorbing Boundary Conditions (ABC) well known from radar calculations, but realization for AW is more complicated and require eigenvector analysis. Now using $\Psi(x)$ we can calculate the wave part of the energy flux
\Eqref{wave_energy_flux} and the wave part of the momentum flux
\be
\tilde\Pi_\mathrm{wave}(x)\equiv
\frac{1}{4\mu_0}\left|\,\hat b(x) \right|^2.
\ee
This section is written in dimensional variables,
but all equations can be easily converted in dimensionless variables
as is done in the next sub-sub-section.
\subsubsection{Dimensionless wave variables, convenient for numerical calculations}
\begin{figure}[t!]
	\begin{center}
	\subfloat[Amplitude profile of the velocity of AW.]{\label{fig:uWave}
	  \includegraphics[width=0.45\textwidth]{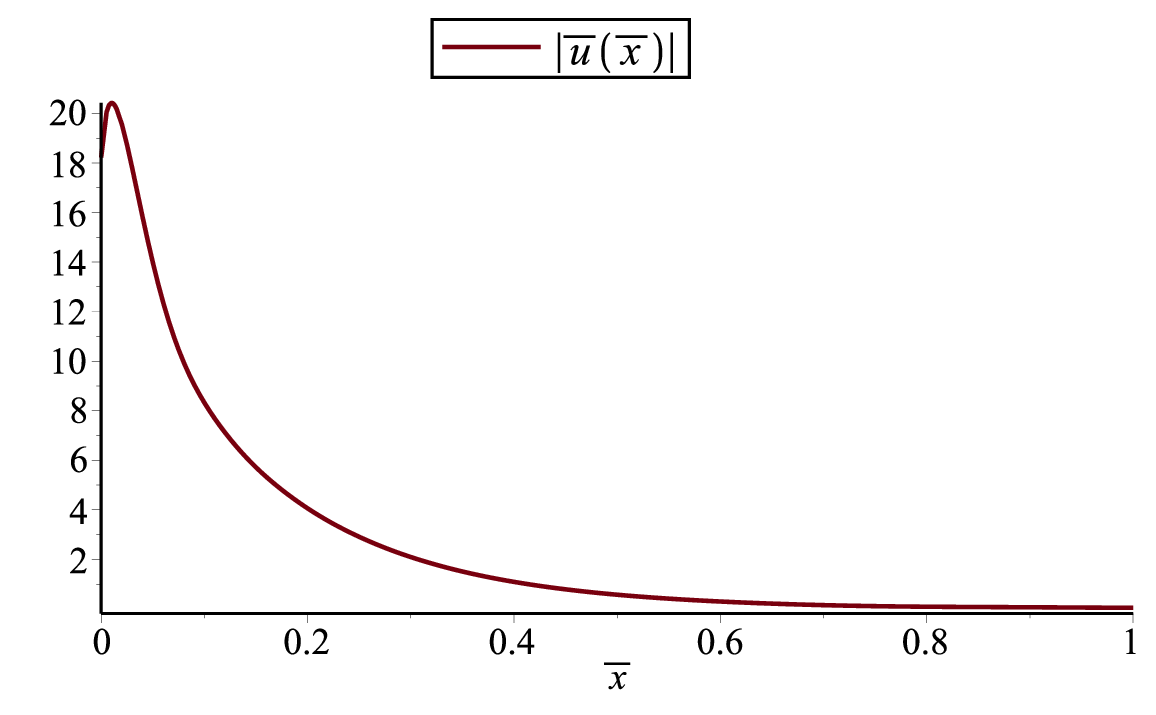}}\\
	\subfloat[Energy flux of AW absorbed by fully ionized Hydrogen plasma according to \Eqref{Qwave}.]{\label{fig:Q}
	  \includegraphics[width=0.45\textwidth]{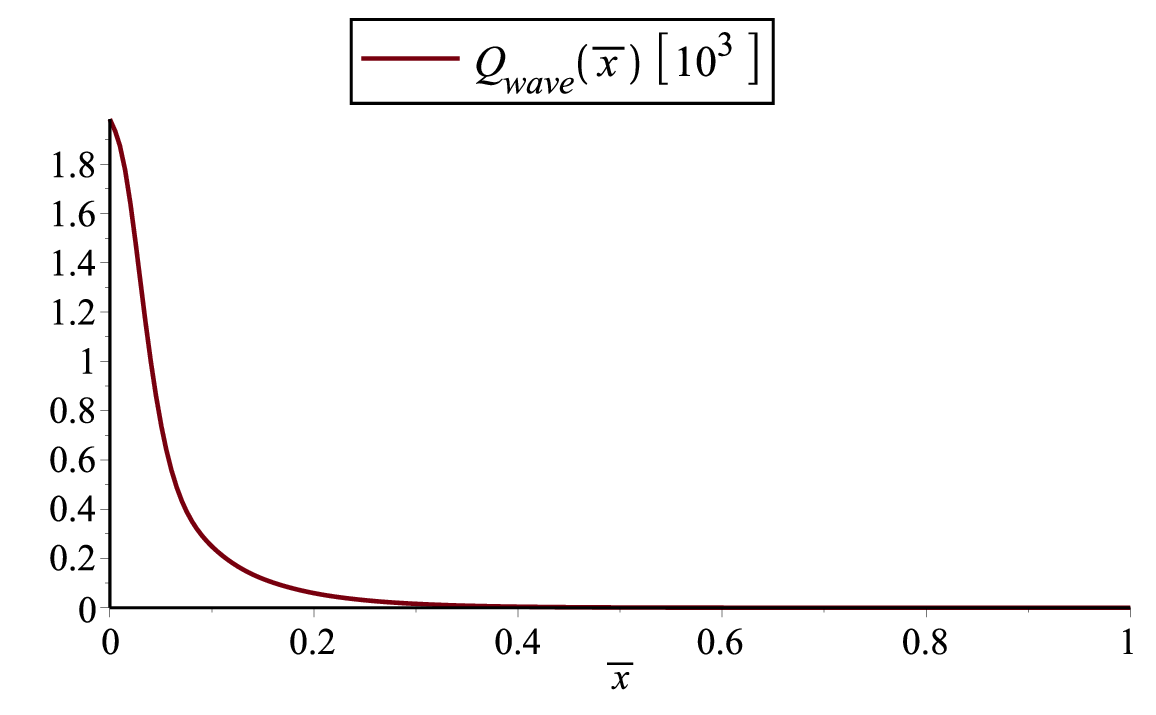}}\\
	\subfloat[Profile of AW momentum flux according to \Eqref{Pwave}.]{\label{fig:P}
	  \includegraphics[width=0.45\textwidth]{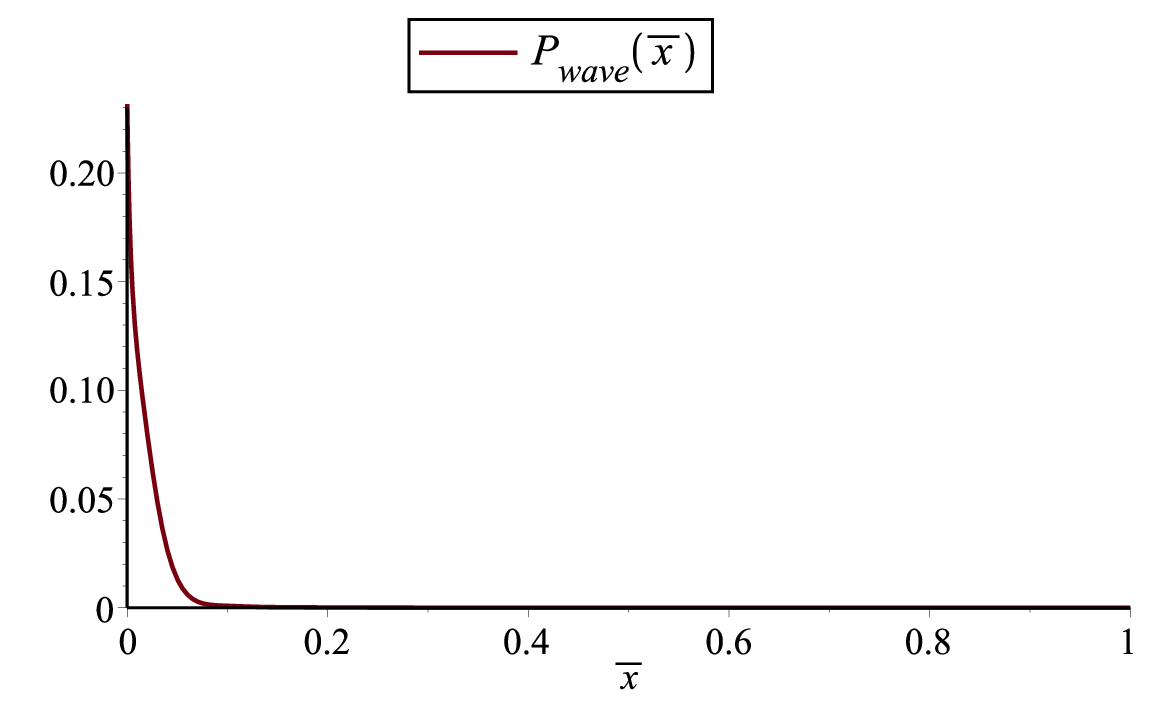}}
	\end{center}
	\caption{Profiles for AW of velocity (a), energy flux (b) and momentum flux (c).}
	\label{fig:wavefluxes}
\end{figure}

Using mechanical units for length $l$, velocity $U_0$
and density $\rho_0$ we can convert all equations in dimensionless form.
The formulae remain almost the same and we wish to mention only the differences.
Introducing dimensionless density
\be
\overline{\rho}(x)= \rho(x)/\rho_0=1/\overline U(\overline x)
\ee
and wave energy flux
\be
\mathcal{Q}_\mathrm{wave}(0)=\frac{q_\mathrm{wave}(0)}{\rho_0U_0^3}=(1-\mathcal R)\left|A_\mathrm{wave}\right|^2, %
\ee
we have dimensionless matrices
\be \mathsf{\overline{M}}\!=\!\!\left(\begin{array}{ccc}0&0&-\overline\nu \overline U
\\0&\overline\nu(-\mathrm{i}\overline\omega\!+\overline W) &-\overline \nu
\\\mathrm{i}\overline\omega \overline U&-\overline V_\mathrm A^2(-\mathrm{i}\overline\omega\!+\overline W)&
\left(\overline V_\mathrm A^2\!-\!\overline U^2\right)\!
+\!\overline U^2\mathrm{d}_{\overline x}\overline\eta\end{array}\right)\!,
\ee
and
\bear
\label{wave_energy_flux_dimensionless}
&&\overline g\equiv\frac14\left(\begin{array}{ccc}1&-a^2&- \overline\eta(\overline x)\\
-a^2&2a^2\overline U(\overline x)&0\\
-\overline\eta(\overline x)&0&0
\end{array}\right),\\
&&\overline V_{\mathrm A}^2(\overline x)=a^2 \overline U(\overline x), \quad \overline V_{\mathrm A}^2(0)=a^2 \overline U(0)=a^2\!.
\eear
For the dimensionless energy flux we have (Fig. \ref{fig:Q})
\bear
\label{Qwave}
\mathcal{Q}_\mathrm{wave}(\overline x)\!\!&=&\!\!\frac{1}{4}\left|\hat{\overline{u}}\right|^2
    +\frac{a^2}{2} \left(\!\overline U\left|\hat {\overline{b}} \right|^2
    \!-\mathrm{Re}(\hat {\overline{b}}^*\hat{\overline{u}})\!\right)
\\&&\nn
-\frac12 \overline\eta\, \mathrm{Re}\left(\hat{\overline{u}}^* \hat{\overline{w}}\right)
=\overline\Psi^\dagger\overline g\,\overline\Psi,\nn
\eear
\noindent
where (Fig. \ref{fig:uWave})
\be
\hat{\overline{u}}=\frac{\hat{u}}{U_0},\quad
\hat{\overline{w}}=\frac{l\mathrm{d}_x \hat{u}}{U_0},\quad
\overline\omega=\frac{l\omega}{U_0}.
\ee
Then for the dimensional wave energy flux we have
\be
q_\mathrm{wave}(x)=\rho_0U_0^3 \, \mathcal{Q}_\mathrm{wave}(\overline x),
\ee
and analogously for the momentum flux (Fig. \ref{fig:P}) of the wave
\be
\label{Pwave}
\Pi_\mathrm{wave}(x)=\rho_0U_0^2\,\mathcal{P}_\mathrm{wave}(\overline x),\quad
\mathcal{P}_\mathrm{wave}=\frac14\left|\hat{\overline b}\right|^2.
\ee
The point in Fig. \ref{fig:P} after which the momentum flux $\mathcal{P}_{\mathrm{wave}}$ remains constant is where it is completely dissipated.
In the next section we will consider all parts of the energy and momentum fluxes.
%%%%%%%%%%%%%%%%%%%%%%%%%%%%%%%%%%%%%%%%%%%%%%
\subsection{Total wave fluxes}
The total energy and momentum fluxes are integrals over all frequencies. From \Eqref{Qwave} we have

\be
\mathcal{Q}_\mathrm{wave, \overline\omega}(\overline x)=\Psi^\dagger_{\overline\omega}\,\overline g\,\Psi_{\overline\omega}=\mathcal{W}_{\overline\omega}\,\psi^\dagger_{\overline\omega}(\overline x)\,\overline g(\overline x)\,\psi_{\overline\omega}(\overline x),
\label{Q_spectr}
\ee
where $\mathcal{W}_{\overline\omega}$ is the spectral density of the waves. We can construct the total energy flux of the waves as
\bear
	\mathcal{Q}^{\mathrm{total}}_{\mathrm{wave}}(\overline x)&=&\int_{\overline\omega=0}^{\infty}\mathcal{W}_{\overline\omega}\,\psi^\dagger_{\overline\omega}(\overline x)\,\overline g(\overline x)\,\psi_{\overline\omega}(\overline x)\frac{\mathrm{d}\overline\omega}{2\pi}\label{qwave}\label{Qmulti}\nn\\
	&=&\sum_{\overline\omega>0}\mathcal{W}_{\overline\omega}\,\psi^\dagger_{\overline\omega}(\overline x)\,\overline g(\overline x)\,\psi_{\overline\omega}(\overline x),\\
	\mathcal{Q}^{\mathrm{total}}_{\mathrm{wave}}(0)\equiv\mathcal{Q}_0&=&\sum_{\overline\omega>0}\mathcal{W}_{\overline\omega}\,\psi^\dagger_{\overline\omega}(0)\,\overline g(0)\,\psi_{\overline\omega}(0).
\eear
Observational data gives a power law dependence of the spectral density of AW in the solar corona. One can suppose that the spectral density of the waves coming from the chromosphere has the same power law dependence, $i.\ e.$ $\mathcal{W}_{\overline\omega}=C/\overline\omega^{\alpha}$. $\alpha$ is between $1.5$ and $2$: $\frac{3}{2}$ \cite{Tomczyk:09}, $1.59$ \cite{Chitta:12}, $2$ \cite{Burlaga:87}. $C$ is the unknown parameter of the theory, which we vary for fixed $\xi$ in order to reproduce the temperature increase in the transition layer. Note that here we have used the dimensionless frequency. If we want to use the dimensional one, then the parameter $C$ also has to become dimensional. If we know the initial total energy flux of the waves, we can calculate the spectral density as
\be
\mathcal{W}_{\overline\omega}=Q_0/\sum_{\overline\omega>0}\psi^\dagger_{\overline\omega}(0)\,\overline g(0)\,\psi_{\overline\omega}(0).
\ee
Analogously to \Eqref{Qmulti}, the total momentum flux is calculated from \Eqref{Pwave} as
\bear
\label{Pmulti}
\mathcal{P}_\mathrm{wave}^\mathrm{total}(\overline x)&=&\sum_{\overline\omega>0}\mathcal{W}_{\overline\omega}\,\mathcal{P}_\mathrm{wave, \overline\omega}(\overline x)\\
&=&\int_{\overline\omega=0}^{\infty}\mathcal{W}_{\overline\omega}\,\mathcal{P}_\mathrm{wave, \overline\omega}
(\overline x)\frac{\mathrm{d}\overline\omega}{2\pi}.\nn
\eear
In order to simulate plasma heating by AW with power law spectral density, in the work by Topchiyska \cite{Topchiyska:13} an illustration is given with 8 AW with different frequencies. Wave propagation can be easily seen for moderate of $T(l)/T_0=3$. In order to concentrate our attention on a realistic temperature increase $T(l)/T_0=20$ in the present work we take into account only one wave with frequency $300$~Hz. No doubt waves in the Hz range do not exist in the solar corona because they are absorbed during the heating, but we wish to emphasize the importance of Hz range waves in the solar photosphere, which are not observable at the moment.

%%%%%%%%%%%%%%%%%%%%%%%%%%%%%%%
\subsection{Mass, energy and momentum fluxes}

In the one-dimensional model which we analyze the conservation laws \Eqref{conslaws}, \Eqref{conslaws2} and \Eqref{conslaws3} are converted in three integrals of our dynamic system describing the mass $j=\rho_0U_0\overline{j},$ energy $q=\rho_0U_0^3\mathcal{Q},$ and momentum $\Pi=\rho_0U_0^2\mathcal{P}$ fluxes
\bear
	\overline{j}&=&\overline U \overline \rho= 1,\\	\label{dimensionless_total_energy_flux}
	\mathcal{Q}&=&\frac12 \overline{U}^2+c_p\Theta_0\overline{T}\nn\\ \label{eq:Qtotal}
	&&-\overline\varkappa \, \Theta_0d_{\overline x}\overline{T}-\left(\frac43\overline\eta+\overline\zeta\right)\overline U\,d_{\overline x}\overline U\nn\\
	&&+\sum_{\overline\omega>0}\mathcal{W}_{\overline\omega}\left(\frac{1}{4}\left|\hat{\overline{u}}_{\overline\omega}\right|^2+\frac{a^2}{2}\left(\overline{U}\left|\hat {\overline{b}}_{\overline\omega} \right|^2-\mathrm{Re}\left(\hat{\overline{b}}_{\overline\omega}^{*}\hat{\overline{u}}_{\overline\omega}\right)\right)\right)\nn\\
	&&-\sum_{\overline\omega>0}\mathcal{W}_{\overline\omega}\frac12 \overline\eta\,\mathrm{Re}\left(\hat{\overline{u}}_{\overline\omega}^*\hat{\overline{w}}_{\overline\omega}\right)=\mathrm{const},\\
	\mathcal{P}&=&\overline{U}+ \frac{\Theta_0\overline{T}}{\overline{U}}-\left(\frac43\overline\eta+\overline\zeta\right)d_{\overline x}\overline{U}\\
	&&+\sum_{\overline\omega>0}\mathcal{W}_{\overline\omega}\frac{1}{4}\left|\,\hat {\overline{b}}_{\overline\omega} \right|^2=\mathrm{const}.\nn\label{dimensionless_total_momentum_flux}
\eear
Here we can recognize the energy flux of an ideal inviscid gas
\be
	\mathcal{Q}_{\mathrm{wind}}^{\,\mathrm{ideal}}=\frac12 \overline{U}^2+c_p\Theta_0\overline{T},\qquad \Theta=\Theta_0\overline T,
	\label{eq:Qidealwind}
\ee
dissipative energy flux of the wind related to heat conductivity and viscosity
\be
	\mathcal{Q}_{\mathrm{wind}}^{\,\mathrm{diss}}=-\overline\varkappa \, \Theta_0d_{\overline x}\overline{T}-\left(\frac43\overline\eta+\overline\zeta\right)\overline U\,d_{\overline x}\overline{U},
	\label{eq:Qdisswind}
\ee
the non-absorptive part of the wave energy flux
\[
\mathcal{Q}_{\mathrm{wave}}^{\mathrm{ideal}}=\sum_{\overline\omega>0}\mathcal{W}_{\overline\omega}\left(\frac{1}{4}\left|\hat{\overline u}_{\overline\omega}\right|^2+\frac{a^2}{2}\left(\overline{U}\left|\hat {\overline{b}}_{\overline\omega} \right|^2-\mathrm{Re}\left({\hat{\overline{b}}_{\overline\omega}}^{*}\hat{\overline u}_{\overline\omega}\right)\right)\right),
\]
\vspace{-6mm}
\be
\label{eq:Qidealwave}
\ee
and the absorptive part of the wave energy flux
\be
	\mathcal{Q}_{\mathrm{wave}}^{\mathrm{\,diss}}=-\sum_{\overline\omega>0}\mathcal{W}_{\overline\omega}\frac12 \overline\eta\,\mathrm{Re}\!\left(\hat{\overline u}_{\overline\omega}^*\hat{\overline{w}}_{\overline\omega}\right)
	\label{eq:Qdisswave}
\ee
proportional to the viscosity. Analogously for the momentum flux we have:
\bear
	\mathcal{P}_{\mathrm{wind}}^{\mathrm{\,ideal}}&=&\overline{U}+\frac {\Theta_0\overline{T}}{\overline{U}},\\
	\mathcal{P}_{\mathrm{wind}}^{\mathrm{\,diss}}&=&\!-\!\left(\frac43\overline\eta+\overline\zeta\right)d_{\overline x}\overline{U},\\
	\mathcal{P}_{\mathrm{wave}}^{\mathrm{\,ideal}}&=&\sum_{\overline\omega>0}\mathcal{W}_{\overline\omega}\frac{1}{4}\left|\,\hat {\overline{b}}_{\overline\omega} \right|^2,\\
	\mathcal{P}_{\mathrm{wave}}^{\mathrm{\,diss}}&=&0\label{Pdisswave}
\eear
for the transversal AW. As a rule the dissipative fluxes are against the non-dissipative ones. One can introduce non-ideal wind energy flux
\bear
	\label{tilde_Q}
	&&\tilde{\mathcal{Q}}\equiv\mathcal{Q}_\mathrm{wind}^\mathrm{nonideal}=\mathcal{Q}_\mathrm{wind}^\mathrm{diss}+ \mathcal{Q}^\mathrm{total}_\mathrm{wave}=\mathcal{Q}-\mathcal{Q}_\mathrm{wind}^\mathrm{ideal}=\\
	&&-\overline\varkappa d_{\overline x}\overline{T}-\left(\frac43\overline\eta+\overline\zeta\right)\overline U d_{\overline x}\overline{U}-\sum_{\overline\omega>0}\mathcal{W}_{\overline\omega}\left(\frac12 \overline\eta\mathrm{Re}\left(\hat{\overline{u}}_{\overline\omega}^*\hat{\overline{w}}_{\overline\omega}\right)\right)\nn\\
	&&+\sum_{\overline\omega>0}\mathcal{W}_{\overline\omega}\left(\frac{1}{4}\left|\hat{\overline{u}}_{\overline\omega}\right|^2+\frac{a^2}{2}\left(\overline{U}\left|\hat {\overline{b}}_{\overline\omega} \right|^2-\mathrm{Re}\left(\hat{\overline{b}}_{\overline\omega}^{*}\hat{\overline{u}}_{\overline\omega}\right)\right)\right),\nn
\eear
\be
	\mathcal{Q}^\mathrm{total}_\mathrm{wave}=\mathcal{Q}_\mathrm{wave}^\mathrm{ideal}+ \mathcal{Q}_\mathrm{wave}^\mathrm{diss}.
\ee
The non-ideal wind momentum flux is
\bear
	\label{tilde_Pi}
	\tilde{\mathcal{P}}&\equiv&\mathcal{P}_\mathrm{wind}^\mathrm{nonideal}=\mathcal{P}_\mathrm{wind}^\mathrm{diss}+ \mathcal{P}^\mathrm{total}_\mathrm{wave}=\mathcal{P}-\mathcal{P}_\mathrm{wind}^\mathrm{ideal}\\\nn
	&=&-\left(\frac43\overline\eta+\overline\zeta\right)d_{\overline x}\overline{U}+\sum_{\overline\omega>0}\mathcal{W}_{\overline\omega}\frac{1}{4}\left|\,\hat {\overline{b}}_{\overline\omega} \right|^2 ,\\
	&&\mathcal{P}^\mathrm{total}_\mathrm{wave}=\mathcal{P}_\mathrm{wave}^\mathrm{\,ideal}+ \mathcal{P}_\mathrm{wave}^\mathrm{diss}.\label{P_wave}
\eear
Then according to \Eqref{sigma-tau} we have
\bear
	\chi(\overline x)&\equiv&\left.\tilde{\mathcal{Q}}\right|_{\overline x}^0=\left.\mathcal{Q}_{\mathrm{wind}}^{\,\mathrm{ideal}}\right|^{\overline x}_0,\\
	\tau(\overline x)&\equiv&\left.\tilde{\mathcal{P}}\right|_{\overline x}^0=\left.\mathcal{P}_{\mathrm{wind}}^{\,\mathrm{ideal}}\right|^{\overline x}_0.
	\label{last_chi_tau}
\eear
This analysis finally reveals the usefulness of dimensionless variables. The dimensional momentum flux $\tau$ and energy flux $\chi$ participate in the important analytical expressions for the wind variables \Eqref{U(x)} and \Eqref{T(x)}.
%%%%%%%%%%%%%%%%%%%%%%%%%%%%%%%%%%%%%%%%%%%%%%%%%%
\section{Self-consistent procedure and results}\label{sec:SCP}
First we fix the boundary condition, the temperature $T_0$ and proton density $n_p(0)$ for $x=0.$ For these parameters we calculate density $\rho_0$, Debye radius length $r_\mathrm{D}(0)$, Coulomb logarithm $\Lambda_0$, viscosity $\eta_0,$ heat conductivity $\varkappa_0$, Ohmic resistivity $\varrho_0,$ and sound speed $c_\mathrm{s}(0).$ Initial velocity of the wind is better to be parameterized by the dimensionless parameter $s\gg1$, i.e. $U_0=c_{\mathrm{s}}(0)/s.$ Analogously plasma beta parameter $\beta_0$ determines the Alfv\'en speed at $V_\mathrm{A}(0)= \sqrt{\frac{\gamma}{2\beta}}c_\mathrm{s}(0).$ Let us also fix the maximal frequency for which we will consider plasma waves $\omega$
and calculate the absorption rate of the energy density of AW $2k^{\prime\prime}_{\mathrm{R}}(0).$ One can choose the interval of the solution of MHD equations to be much larger than the AW mean free path $1/2k^{\prime\prime}_{\mathrm{R}}(0),$ for example
\be
	l=\frac{10}{2k^{\prime\prime}_{\mathrm{R}}(0)}=\frac{10\Va^3(0)}{\nu_\mathrm{k}(0)\omega^2}.
\ee
We can now explain why we have implicitly neglected the gravitational effects. The reason is that our spatial scale $l$ is much smaller than the barometric scale $H=k_{\mathrm{B}}T/\left<m\right> g_{\!_{\odot}}$, $l\ll H$. Here we have used the standard notation for the Sun's surface gravity $g_{\!_{\odot}}$. Having units for length $l$, velocity $U_0$ and density $\rho_0$ we can calculate dimensionless  variables at $\overline x=0:$ $\overline\varkappa_0,$ $\overline\eta_0,$ and $\Theta_0.$

\begin{figure}[t!]
  \begin{center}
  \subfloat[Space dependence of the temperature, calculated by our self-consistent MHD procedure.]{\label{fig:Tx}\includegraphics[width=0.42\textwidth]{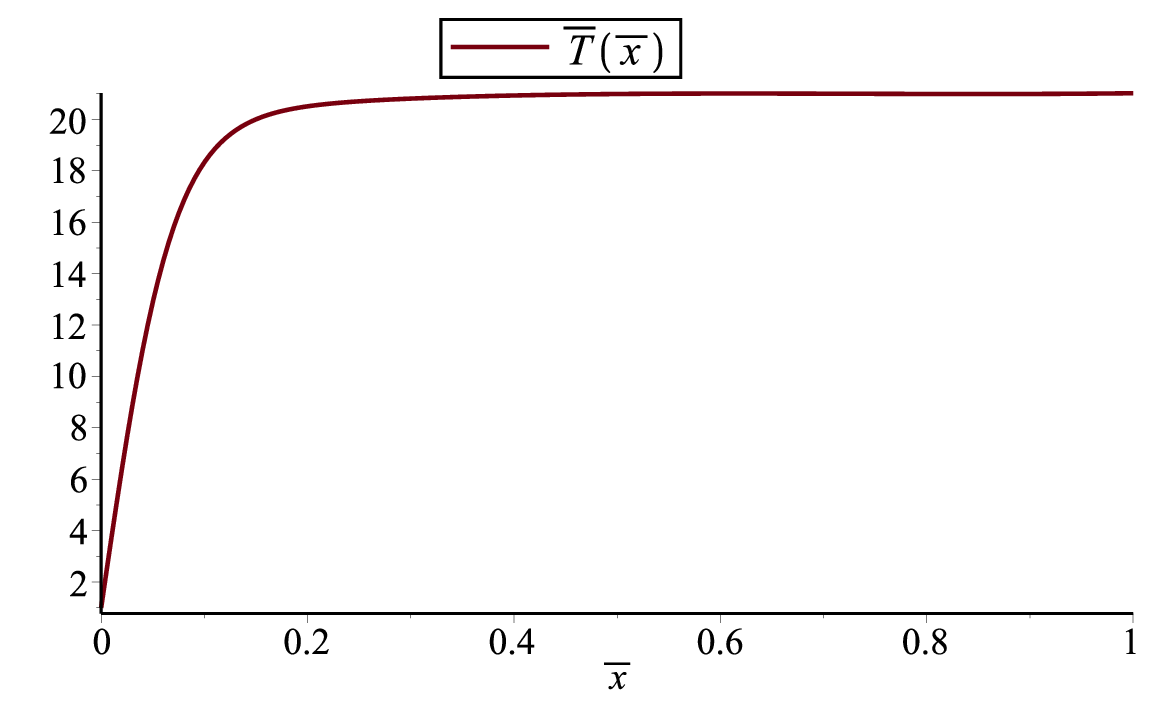}}\\
  \subfloat[Relative difference between consecutive temperature profiles $\overline T_\mathrm{old}(x)$ and $\overline T(x)$. The smallness of this difference is an empirical criterion for the convergence of the calculation.]{\label{fig:TSolErr}\includegraphics[width=0.42\textwidth]{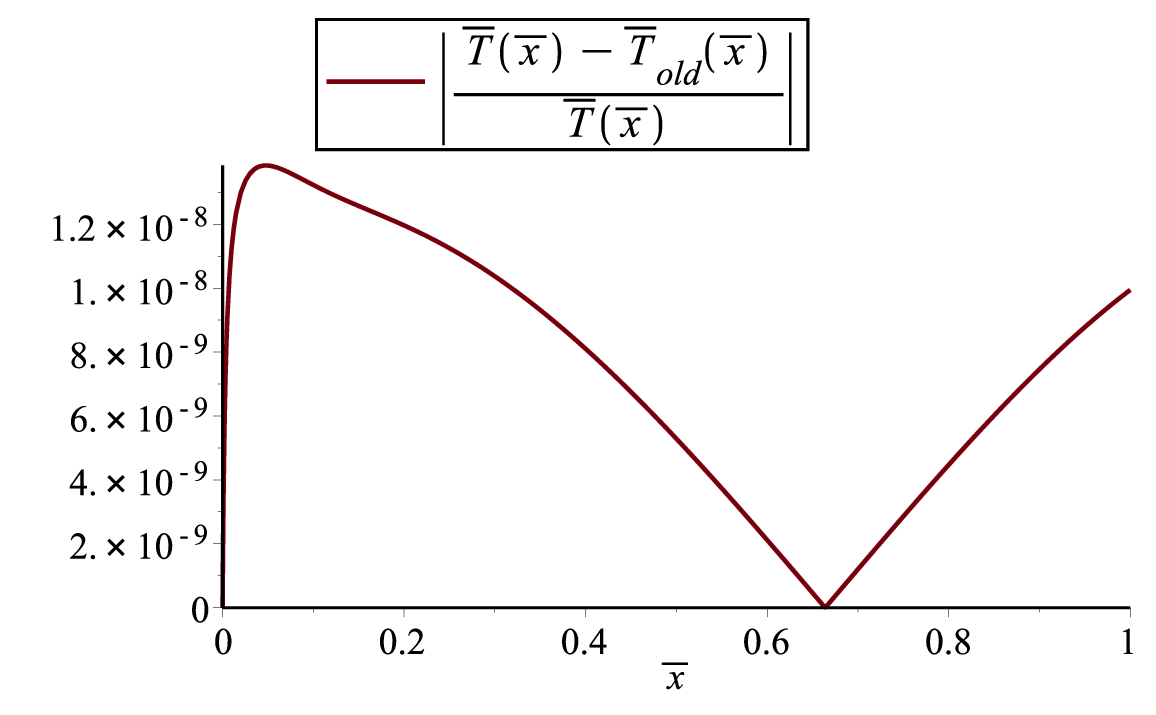}}
  \end{center}
  \caption{A twenty times increase of the plasma temperature by absorption of AW: (a) temperature profile; (b) relative difference for the self-consistent MHD calculations, as described in section \ref{sec:SCP}.}
  \label{fig:T}
\end{figure}

\begin{figure}[t!]
  \begin{center}
  \subfloat[Space dependence of the velocity, calculated by our self-consistent MHD procedure.]{\label{fig:Ux}\includegraphics[width=0.42\textwidth]{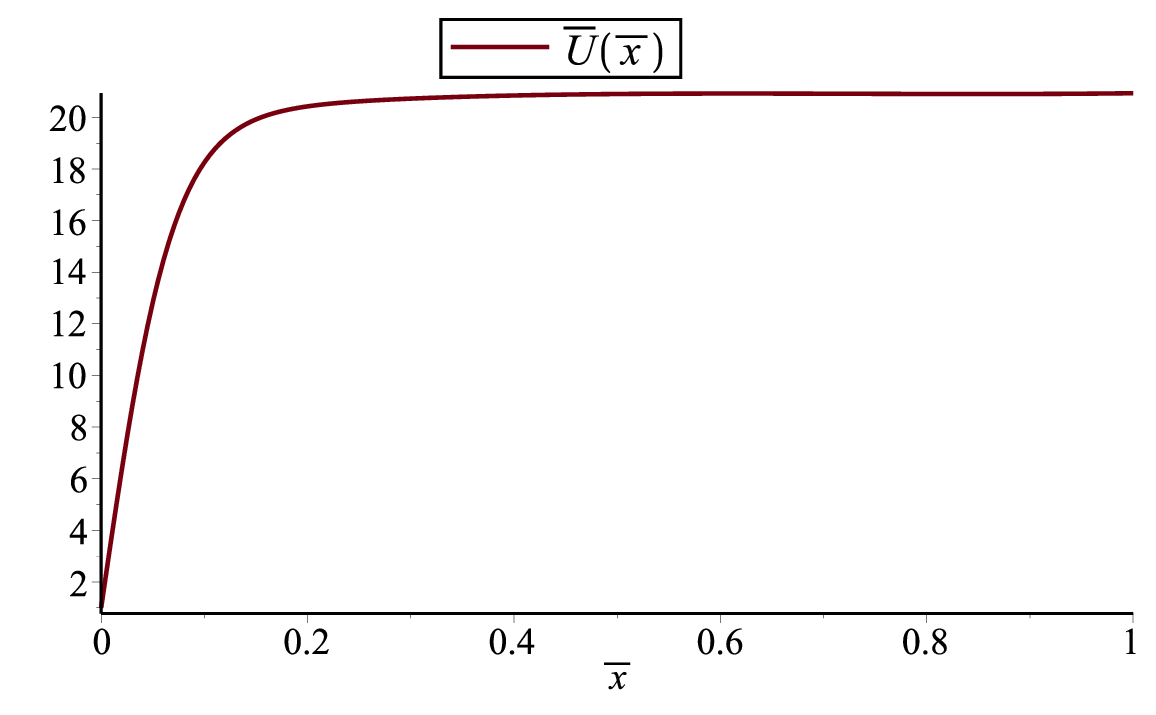}}\\
  \subfloat[Relative difference between consecutive velocity profiles $\overline U_\mathrm{old}(x)$ and $\overline U(x)$.]{\label{fig:USolErr}\includegraphics[width=0.42\textwidth]{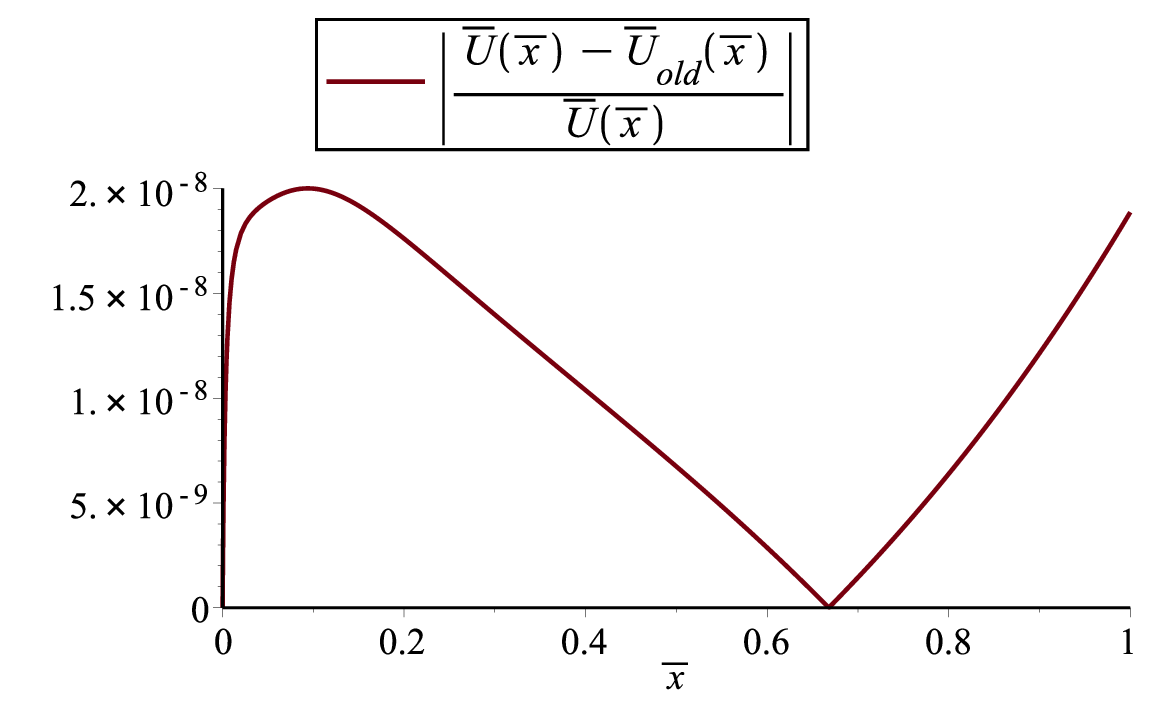}}
  \end{center}
  \caption{A twenty times increase of the plasma velocity by absorption of AW: (a) velocity profile; (b) relative difference for the self-consistent MHD calculations, as described in section \ref{sec:SCP}. The figures are analogous to Fig.~\ref{fig:T}.}
  \label{fig:U}
\end{figure}

The input parameters of the program are $T_0,$ $n_{\mathrm{tot}}(0),$ $\beta_0,$ $\omega,$ and $s$
which parameterizes $j=\left<m\right>n_{\mathrm{tot}}(0)U_0.$
We calculate $l,$ $a,$ $\overline \eta_0=\overline \nu_0,$ $\overline \varkappa_0$ and choose some $A_\mathrm{wave}$
which finally determines increasing of the temperature $\overline{T}(1)=T(l)/T(0).$

In our self-consistent calculation we use the non-linear fit $\overline{T}(\overline{x})=1+\left(\sum_{n=0}^3a_n\overline{x}^n\right)\tanh(b_1\overline{x})$ for the numerically calculated profile of the temperature $\overline{T}(\overline{x})$ as well as for the velocity $\overline{U}(\overline{x})$. In order to accelerate the convergence for the initial approximation we use $a_0=20$, $a_1=a_2=a_3=0$, $b_1=10$.
Let us explain in detail the successive approximations.
%%%%%%%%%%%%%%%%%

\begin{enumerate}
\item At fixed wind profiles $\overline T(\overline{x})$ and $\overline{U}(\overline{x})$ we calculate $\Lambda=\Lambda_0+\frac32\ln\overline T +\frac12\ln\overline U,$ $\overline \eta,$
$\overline \varkappa,$ $\mathrm d_{\overline{x}}\overline\eta,$ and dimensionless matrices
$\overline{\mathsf{M}}$ and $\overline g.$ Then we have to solve the wave equation and to renormalize
the solution with some fixed dimensionless energy flux for the R-mode $\mathcal{Q}_\mathrm{wave}(0).$
\item Using so obtained wave variables $\Psi$ we have to solve equations \Eqref{q/j} and \Eqref{Pi/j} to find $\overline U(\overline x)$ and $\overline\Theta(\overline x)$, and respectively $\overline T(\overline x)$ from \Eqref{dlesswind}.
%the ordinary differential equations for
%$\mathrm{d}_{\overline{x}}\overline{F}$ and $\mathrm{d}_{\overline{x}}\overline{W},$
%with boundary conditions, for example, $\overline{F}(0)=0$ and $\overline{W}(0)=0.$
The variables $\overline \eta,$
$\overline \varkappa,$ $\mathrm d_{\overline{x}}\overline\eta$, and
$\mathrm d_{\overline{x}}\overline\varkappa,$ which participate in
the coefficients have to be calculated simultaneously.
\item Having solved the equations for the wind variables and formerly the equations for the wave variables
we can calculate the total energy and momentum flux. If the maximal relative difference between two successive temperature profiles is larger than some predetermined value we go to step 1 and repeat the procedure.
\end{enumerate}
The width of the transition layer $\lambda$, defined by
\be
\frac l\lambda=\max_{\overline x} \left|\mathrm d_{\overline{x}}\,\ln\overline T \right|\simeq\frac{1}{b_1},
\ee
and the increasing of the temperature $\overline T(1)$ are functions of the wave energy flux coming from the chromosphere $q_{\mathrm{wave}}(0)$. The wave amplitude $A_\mathrm{wave}$ is determined in a way that ensures the desired temperature increase $\overline T(1;A_\mathrm{wave})=\overline{T}(l)/\overline{T}(0)$ in the end of the space interval. In our numerical illustration we used one wave which corresponds to $\delta$-like spectral density of AW.
In general, at fixed temperature increase and chosen spectral density of the
AW the MHD theory has no more free parameters. All that is left is to compare the
calculations with other models and observational data for heating of the solar corona and launching of the solar wind.

We have demonstrated that a $21$ times increase in the temperature is possible by setting $T_0=6000~\mathrm{K}$, $n_{\mathrm{tot}}(0)=4\cdot 10^{17}~\mathrm{m}^{-3}$, $\beta_0=1$, $\omega=2\pi\,10~\mathrm{rad}/\mathrm{s}$, $s=137$. The final dimensionless temperature profile and relative difference are shown in Fig.~\ref{fig:T}. In Fig.~\ref{fig:Tx} one can easily see the transition between strong absorption of AW and wind with evanescent AW amplitude wave and almost horizontal temperature profile. We have shown only the beginning of the interval, where the AW are absorbed. The dimensionless wind velocity profile and relative difference are shown in Fig.~\ref{fig:U}.

\begin{figure}[h!]
\begin{center}
\includegraphics[width=0.42\textwidth]{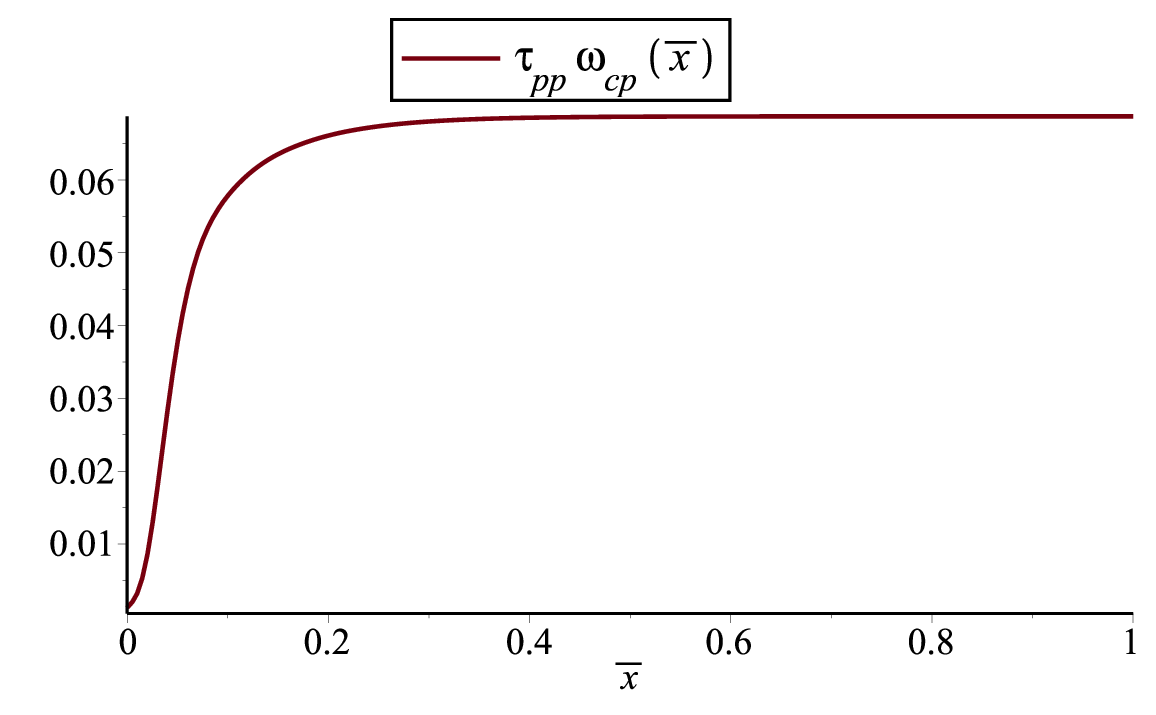}
\caption{Spatial dependence of the applicability criterion.}
\label{fig:crit}
\end{center}
\end{figure}

Few more comments are worthwhile. The applicability criterion (\ref{eq:crit}) is fulfilled as can be seen in Fig.~\ref{fig:crit}.
The stiffness ratio in the beginning of the interval is above 2000 which is very high and confirms that we are dealing with a stiff system, thus making the problem very hard from a numerical point of view. After we have illustrated the principle of the heating mechanism, we should note that perturbative accessories could be added to the MHD equations, i.e. radiative losses, chemical content, and Ohmic heating. However, this will not change in any way the temperature profile. Finally, a realistic application of the MHD heating of the solar corona includes realistic spectral density of AW, coming from the chromosphere, as a boundary condition for the transition region.

%%%%%%%%%%%%%%%%%%%%%%
\section{Conclusion}
%%%%%%%%%%%%%%%%%%%%%%
\subsection{Discussions}
In spite that iron was suspicious from the very beginning the
problem of Coronium was a 70 year standing mystery until unambiguous
identification as Fe$^{13+}$ by Grotrian and Edlen in 1939. The same
70 year time quantum was repeated. In 1947 \alf \cite{Alfven:47} advocated the idea that
absorbtion of AW is the mechanism of heating of solar corona.
Unfortunately the idea by Swedish iconoclast \cite{Dessler:70} was
never realized  in original form: what can be calculated, what is
measured, what is explained and what is predicted. That is why there
is a calamity of ideas still on the arena, for a contemporary review
see the SOHO proceedings (\cite{SOHO:04}). From qualitative point of
view the narrow width of the transition layer
$w=\mathrm{min}\,\left|\mathrm{d} x/ \mathrm{d} \ln T(x)\right|$ is
the main property which should be compared against the predictions
of other scenarios. For example, in order for the nanoflare hypothesis to be
vindicated \cite{Day:09} such reconnections are needed to explain the
narrow width of the transition layer at the same boundary conditions
of wind velocity and temperature. Moreover electric fields of the
reconnections heats mainly the electron component of the plasma. How
then proton temperature in the corona is higher?  Launching of
Hinode gave a lot of hints for the existence of AW in the corona
\cite{Pontieu:07}, see also \cite{Jess:09}. However most of the
those research was in UV region when high frequency AW which heated
are already absorbed. All observations are for low frequency (mHz
range) AW for which hot corona is transparent. The best can be done
is the extract low frequency behavior of the spectral density of AW
and to extrapolate to higher frequencies responsible for heating. So
observed AW are irrelevant for the heating. In order to identify AW
responsible for the heating it is necessary to investigate high
frequency (1 Hz range) AW in the cold chromosphere using optical not
UV spectral lines. We are unaware whether such type of experiments
are planned. One of the purposes of the present work
is to focus the attention of experimentalists on the 1~Hz range AW
in the chromosphere, which we predict on the basis of our MHD
analysis. For such purposes we suggest to be paid special attention to the behavior of oscillations \cite{Kolobov:15} and sunspot waves \cite{Chandra:15} above active regions. Another possibility is provided by Doppler tomography \cite{Marsh:05} of H$\alpha$ or Ca lines. Doppler tomography was successfully used for investigation of rotating objects, such as  accretion disks
\cite{Marsh:88} and solar tornados \cite{tornado}. Here we wish to
mark also the Doppler tomography by Coronal Multi-channel
Polarimeter build by Tomczyk \cite{Tomczyk:09}. For investigation of
AW by Doppler tomography we suggest development of frequency
dependent Doppler tomography operating as a lock-in voltmeter. The
date from every space pixel should be multiplied by $\sin\omega t$
and integrated for many wave periods. Finally one can observe time
averaged distribution of the AW amplitude. Systematic investigation
of such frequency dependent Doppler tomograms will reveal that
Swedish iconoclast \cite{Dessler:70} is again right that AW heat the
solar corona, after another 70 years of dramatic launching of
vast variety of ideas.
\subsection{Plasma heating by AW -- a historical perspective}
What have we learned from the one-dimensional static MHD problem? We have demonstrated that qualitatively predicted self-induced opacity of plasma is an intrinsic property. Absorption of AW causes viscous heating of ions and that is why the proton temperature is higher than the electron one. In this way we have revealed an effective method for ion heating which can be applied to many plasma problems. Actually plasma heating by MHD waves is used in the MIT alcator \cite{Snipes:05}. We suggest however that the toroidal geometry can be replaced by Budker probkotron geometry, in which the energy of the AW will be focused in a narrow jet with a hundred times increased temperature. A de Laval nozzle will be realized by strong magnetic fields. We do believe that this will be an effective method for navigation in the Solar System (cf. \cite{Choueiri:09}). Electric power from a nuclear reactor will create a fast electron-proton jet and this will dramatically decrease the initial mass of the rocket. For large-scale Earth-based installations such a jet of high-temperature deuterium will inject a fresh idea in nuclear fusion physics.

\begin{acknowledgement}
Authors are thankful to Yana Maneva \cite{Araneda:09,Maneva:09,Maneva:10,Maneva:10b} and Martin Stoev
for the collaboration in the early stages of the present research \cite{Mishonov:07} when
the idea of self-induced opacity was advocated and consideration of many 
problems related to physics of solar corona. Fruitful comments and discussions with Tsvetan Sariisky are highly appreciated. 
Authors are also thankful to Eckart Marsch for the hospitality during the conference "From the Heliosphere into the Sun -- Sailing Against the Wind" (http://www.mps.mpg.de/meetings/hcor/) and to Ivan Zhelyazkov, Steven R. Cranmer, Leon Ofman, Jaime Araneda, Plamen Angelov, Ramesh Chandra, Yurii Dumin, and Yana Maneva for the interest and comments. This work was partially supported by Indo-Bulgarian scientific grant \textnumero~CSTC/India 01/7.
\end{acknowledgement}
%%%%%%%%%%%%%%%%%%%%%%%%%%%%%%%%%%%%%%%

\end{document}